# Theoretical Prediction of the Robust Intrinsic Half-Metallicity in Ni$_2$N MXene with Different Types of Surface Terminations

Guo Wang* and Yi Liao

Department of Chemistry, Capital Normal University, Beijing 100048, China

*Email: wangguo@mail.cnu.edu.cn

**Abstract**

Bare and surface-passivated Fe$_2$N, Co$_2$N, and Ni$_2$N MXene were investigated by using density functional theory. Fe$_2$N(OH)$_2$, Fe$_2$NO$_2$, Co$_2$NO$_2$, Ni$_2$NF$_2$, Ni$_2$N(OH)$_2$, and Ni$_2$NO$_2$ are intrinsic half-metals, while other structures have antiferromagnetic ground states. The half-metallicity of Ni$_2$NT$_2$ (T = F, OH, and O) does not depend on the type of surface terminations and should be more realizable in experiments. The energy differences between the ferromagnetic and antiferromagnetic configurations of Ni$_2$NT$_2$ are several hundreds of meV per primitive cell. The Curie temperature should be above room temperature from the point of view of mean field approximation.

## Introduction

MXene, a transition metal carbide or nitride, has received much attention because of its promising applications in catalysts, sensors and energy storage.[1] Besides, the excellent conductivity makes MXene a good candidate for electrodes. Recently, both experimental and theoretical works have indicated that MXenes can be used as Schottky-barrier-free metal contacts to two-dimensional semiconductors, which are essential for integration of field effect transistors.[2,3] The high strength and flexibility make them excellent candidates for flexible electronics.[4,5] Therefore, MXene also has potential applications in electronics.

In traditional electronics, charge plays the role of carrier. However, spin takes this responsibility in spintronics. In spintronic devices, the power consumption is significantly reduced, but special magnetic materials are required. In a half-metal, one spin channel is metallic while the other is insulating or semiconducting.[6] The spin polarization is 100%. Thus, half-metals are extremely suitable for the electrodes in spintronic devices. For MXenes, $Cr_2C$,[7] $Cr_2NO_2$ [8] and $Mn_2CT_2$ (T = F, Cl or OH) [9] were predicted to be intrinsic half-metals while $Ti_2C$ [10] becomes a half-metal under a biaxial strain. With the development of computer industry, electronic devices become progressively smaller with increasing time. Two-dimensional half-metallic MXenes should further reduce the device size in spintronics.

Usually, MXene is obtained by etching the weakly-bonded A group element [11] from its parent MAX phase.[12] The bare MXenes should be difficult to be maintained in the etching process. The $Cr_2NO_2$ [13,14] and $Mn_2CT_2$ (T = F, Cl, or OH) [9] are half-metallic MXenes with surface terminations. Although the passivated structures are more stable than the bare ones, the half-metallicity still depends on their surface terminations. The $Cr_2N(OH)_2$ and $Mn_2CO_2$ are not half-metals.[8,9] Previous experiments have revealed that the synthesized $Ti_3C_2$ MXene has mixed surface terminations of F, OH, and O.[15,16] Although the oxidization reaction from $Cr_2N(OH)_2$ to $Cr_2NO_2$ was calculated to be energetically favorable,[8] the proportion of oxidization may depend on the experimental condition. These affect their applications in low-dimensional spintronics. Interestingly, a double transition metal carbides MXene [17] $Ti_2MnC_2T_2$ (T = F, OH, or O) [18] was predicted to be a robust magnetic semimetal or metal. The magnetism does not depend on the surface terminations. This benefits its real application in spintronics. Since its spin-polarization is not 100%, a half-metallic MXene that does not depend on its surface terminations is highly desirable. Originally, only MAX phases that contain early transition metals are reported. For 3d metals, these elements are Sc Ti V Cr.[12] Recently, $Mn_2GaC$ was successfully synthesized.[19] The MAX phase family is expanding. This gives us new hope for exploring the properties of the MXene family. It is beneficial to theoretically find out whether the unexplored MXenes have exciting properties before they are synthesized. In the present work, electronic and magnetic properties of $Fe_2N$, $Co_2N$, and $Ni_2N$ MXenes are investigated by using density functional theory.

## Models and computational details

The model of $X_2N$ (X = Fe, Co, and Ni) is shown in Figure 1(a). In this hexagonal layered structure, the N atom layer is sandwiched between two metal layers. The passivated $X_2NT_2$ (T = F, OH, and O) were also investigated. Since the functional atoms or groups are above the hollow site of the N atoms are always directed to the metal atoms in the opposite layer,[20,21] they are not shown in Figure 1 for clarity. For ferromagnetic (FM) configurations, only a primitive cell is needed. Three antiferromagnetic configurations were also considered as shown in Figure 1(b)–1(d). In the AFM1 configuration, the spins are parallel in the same metal layer while antiparallel in the opposite layers. For the AFM2 and AFM3 configurations, a 2×1 supercell is needed. A 2×2 supercell yields another two nominal AFM configurations. However, the two configurations do not really exist [8] and will not be studied here.

The geometrical optimization and calculation of electronic properties were carried out by using the Vienna *ab initio* simulation package (VASP).[22] The PBE density functional [23] and projector-augmented wave with an energy cutoff of 400 eV were used. The strongly correlated 3d electrons were treated in the framework of GGA+U method. The difference between the onsite coulomb (U) and exchange parameter (J) was set to 4.0, 3.3 or 6.4 eV for the 3d orbital of Fe, Co, or Ni atoms.[24] A Monkhorst−Pack sampling with 41×41×1 *k*-points was used in the first Brillouin zone of the two-dimensional hexagonal primitive cell. The vacuum layer is about 15 Å thick to avoid possible interaction between artificial images along the non-periodic direction. Since the MXenes studied here have not been synthesized, phonon dispersion was also calculated. This was fulfilled by using density functional perturbation theory with the aid of the Phonopy code.[25] A 4×4 supercell was used in the calculations. Due to the large computational resources required, the number of *k*-points was reduced by half along the periodic directions in the phonon dispersion calculations.

Tests calculations were performed to confirm the selection of the parameters.

Ni$_2$NO$_2$ was taken as an example. The optimized lattice parameter is 2.909, 2.893, or 2.893 with a 21×21×1, 41×41×1, or 61×61×1 *k*-points. The 41×41×1 *k*-point net is dense enough to obtain converged geometry and the corresponding electronic properties. With this *k*-point sampling, the relative energies of the AFM1, and AFM2 configurations are 647 and 532 meV per primitive cell, taking the energy of the FM configuration as the reference. The AFM3 configuration has not been obtained, although much effort has been made. The total magnetic moment is always non-zero (−0.75 μ$_B$), so the AFM3 configuration will not be studied. When the energy cutoff is increased to 600 eV, the relative energies are 684 and 564 meV. This gives the same FM ground state of Ni$_2$NO$_2$. Although the U–J values were carefully tested in the literature, different U–J values were also considered here. The relative energies of the AFM1 configurations are 87, 155, 310, and 833 meV while those of the AFM2 configurations are 80, 157, 285, and 726 meV, when U–J equals to 0, 2, 4, and 8 eV, respectively. The ground states are always the FM configurations. The total magnetic moments are always 1 μ$_B$ with different U–J values. These confirmed the selection of the parameters.

## Results and discussions

The relative energies of the MXenes studied are listed in Table 1. For the bare MXenes, none of the FM configurations is the ground state. On the contrary, some of the FM structures with surface terminations are the ground states. The magnetism should be more achievable with passivated-surfaces. For Fe$_2$NT$_2$ and Co$_2$NT$_2$, only Fe$_2$N(OH)$_2$, Fe$_2$NO$_2$, and Co$_2$NO$_2$ have FM ground states. Their band structures are shown in Figure 2(a)–2(c). In their band structures, several bands go across the Fermi level for majority spin. They are metallic for majority spin while semiconducting for minority spin. The three FM structures are intrinsic half-metals. The band gaps for minority spin are 1.08, 1.32, and 1.58 eV, while the half-metallic band gaps (defined as the energy difference between Fermi level and valence band maximum)[26] are 0.13, 0.72, and 0.29 eV, respectively. It is noted that the half-metallic band gap for

Fe$_2$N(OH)$_2$ is not so large that thermally induced spin flip may occur at finite temperature. The spin-polarization may be less than 100% at room temperature. Moreover, not all the structures are half-metallic for the surface-passivated Fe$_2$NT$_2$ and Co$_2$NT$_2$. The half-metallicity requires special surface terminations. To achieve this, completely removing F atoms is needed. For Co$_2$NT$_2$, an additional step with complete oxidization is required. These hinder their application because the synthesized MXene has mixed surface terminations of F, OH, and O.[15,16]

For the surface-passivated Ni$_2$NT$_2$, the situation is different. It is exciting that all the structures (T = F, OH, and O) have FM ground states, as indicated in Table 1. Their band structures are shown in Figure 2(d), 2(f), and 2(h), respectively. They are all intrinsic half-metals. The half-metallicity is robust and does not depend on the type of surface terminations. The band gaps for minority spin are 3.67, 2.04, and 2.98 eV, while the half-metallic band gaps are 1.05, 0.91, and 0.54 eV, respectively. These half-metallic band gaps are much larger than the thermal energy $k_\text{B}T$ at room temperature, so thermally induced spin flip should be suppressed.

Because of their importance, the band structures of Ni$_2$NT$_2$ were also calculated with the HSE06 hybrid density functional,[27] which can accurately treat the band gaps of solids.[28] The band structures are shown in Figure 2(e), 2(g), and 2(i), respectively. The band shape is similar with the corresponding one calculated with the PBE density functional. However, the band gaps for minority spin increase to 4.57, 3.11, and 3.23 eV, while the half-metallic band gaps increase to 2.03, 1.81, and 1.01 eV, respectively. The half-metallic band gaps are about twice magnitude of the corresponding values calculated with the PBE density functional. They are larger than 1 eV. Furthermore, the band gaps for minority spin are larger than 3 eV. The structures are insulating for minority spin. Therefore, the half-metallicity (100% spin polarization) of Ni$_2$NT$_2$ is more likely to be achieved. Since the results calculated with the HSE06 hybrid density functional and GGA+U method are in qualitative agreement, the results below are all based on the GGA+U method unless explicitly stated.

Phonon dispersions were calculated to confirm the stability of the three Ni$_2$NT$_2$. In

Figure 3(a)–3(c), no non-trivial imaginary frequency was found in the phonon dispersions. Near the Γ points, small imaginary frequencies exist in the three acoustic phonon bands for each $Ni_2NT_2$. Detailed normal coordinates analysis indicates that the three normal modes are translational modes. In each mode, all the atoms move along the same direction and there is no relative movement between different atoms. Because the density functional theory uses numerical integration grids rather than analytical methods, the frequencies of the three translational modes are not exactly zero. Nevertheless, there is no vibrational modes with imaginary frequency. The phonon dispersions confirm the stabilities of the three $Ni_2NT_2$.

As listed in Table 1, only the half-metallicity of $Ni_2NT_2$ does not depend on the type of surface terminations. The total magnetic moments are 3, 3, and 1 $\mu_B$ for $Ni_2NF_2$, $Ni_2N(OH)_2$, and $Ni_2NO_2$, respectively. Detailed analysis indicates that the magnetic moments on each Ni and N atom are 1.60 and -0.35 $\mu_B$ for $Ni_2NF_2$ while the values are 1.60 and -0.34 $\mu_B$ for $Ni_2N(OH)_2$. The magnetic moments on each Ni, N, and O atom are 0.97, -0.41, and -0.27 $\mu_B$ for $Ni_2NO_2$. The magnetization is mainly on the Ni atoms. In order to further investigate the magnetic coupling between Ni atoms, the energy of the AFM1 and AFM2 configurations were calculated based on the FM geometries. Compared with the FM configuration, the relative energies of the AFM1, and AFM2 configurations are 673 and 638 meV per primitive cell for $Ni_2NF_2$. These values are 567 and 572 meV for $Ni_2N(OH)_2$ while 759 and 790 meV for $Ni_2NO_2$. The energy differences are an order of magnitude larger than $k_B T$ ($T$ is room temperature). From the point of view of mean field approximation, the Curie temperature of the three $Ni_2NT_2$ should be much higher than room temperature.

If the nearest intra- and inter-layer magnetic couplings between Ni atoms are considered, the Hamiltonian is written as follows based on the Heisenberg model

$$H = -\sum_{i,j} J_{intra} S_i \cdot S_j - \sum_{k,l} J_{inter} S_k \cdot S_l$$

in which $S$ is the spin on a Ni atom, $J_{intra}$ and $J_{inter}$ are intra- and inter-layer exchange parameters. Based on this model, the following equations are written:

$$E_{FM} = (-6J_{intra} - 3J_{inter})S^2 + E_0$$

$$E_{\text{AFM1}} = (-6J_{\text{intra}}+3J_{\text{inter}})S^2+E_0$$

$$E_{\text{AFM2}} = (2J_{\text{intra}}-J_{\text{inter}})S^2+E_0$$

The exchange parameters were obtained by calculating the energy difference between the FM and AFM configurations. The calculated $J_{\text{intra}}$ equals to 81, 75, or 285 meV while $J_{\text{inter}}$ equals to 175, 148, or 538 meV for $Ni_2NF_2$, $Ni_2N(OH)_2$, or $Ni_2NO_2$, respectively. The higher values for $Ni_2NO_2$ should be partially attributed to the lower magnetic moment. The huge positive exchange parameters indicate that the Ni atoms have both strong intra- and inter-layer FM couplings, and the half-metallicity should survive at room temperature.

## Conclusions

Three $Fe_2N$, $Co_2N$, and $Ni_2N$ MXene as well as the surface-passivated structures were investigated by using density functional theory. All the bare MXenes have AFM ground states, while $Fe_2N(OH)_2$, $Fe_2NO_2$, $Co_2NO_2$, $Ni_2NF_2$, $Ni_2N(OH)_2$, and $Ni_2NO_2$ have FM ground states. All the MXenes with FM ground states are intrinsic half-metals. Especially, the half-metallicity of $Ni_2NT_2$ (T = F, OH, and O) does not depend on the type of surface terminations. Because the synthesized MXene has mixed surface terminations of F, OH, and O, the half-metallicity of $Ni_2NT_2$ should be more realizable in experiments. For $Ni_2NT_2$, the energy difference between the FM and AFM configurations are 438−647 meV per primitive cell. The total magnetic moments are 3, 3, and 1 $\mu_B$ for $Ni_2NF_2$, $Ni_2N(OH)_2$, and $Ni_2NO_2$, respectively. Detailed analysis indicates that the magnetization occurs mainly on the Ni atoms. By calculating the energy difference between the AFM and FM configurations based on the FM geometries, exchange parameters were obtained based on the Heisenberg model. The intra-layer exchange parameters are 81, 75, and 285 meV while the inter-layer ones are 175, 148, and 538 meV for $Ni_2NF_2$, $Ni_2N(OH)_2$, and $Ni_2NO_2$, respectively. The large energy difference between the FM and AFM configurations and the huge exchange parameters indicate that the half-metallicity should survive at room temperature, from the point of view of mean field approximation. The $Ni_2N$

MXene with surface terminations should be a good candidate for two-dimensional spintronics.

Table 1. Relative energies (eV) of the AFM configurations.

|  | AFM1 | AFM2 |
|---|---|---|
| $Fe_2N$ | −199 | −278 |
| $Fe_2NF_2$ | −23 | −405 |
| $Fe_2N(OH)_2$ | 261 | 2132 |
| $Fe_2NO_2$ | 218 | 45 |
| $Co_2N$ | −245 | −223 |
| $Co_2NF_2$ | −150 | −915 |
| $Co_2N(OH)_2$ | −108 | −680 |
| $Co_2NO_2$ | 344 | 330 |
| $Ni_2N$ | −1208 | −1289 |
| $Ni_2NF_2$ | 469 | 438 |
| $Ni_2N(OH)_2$ | 452 | 457 |
| $Ni_2NO_2$ | 647 | 532 |

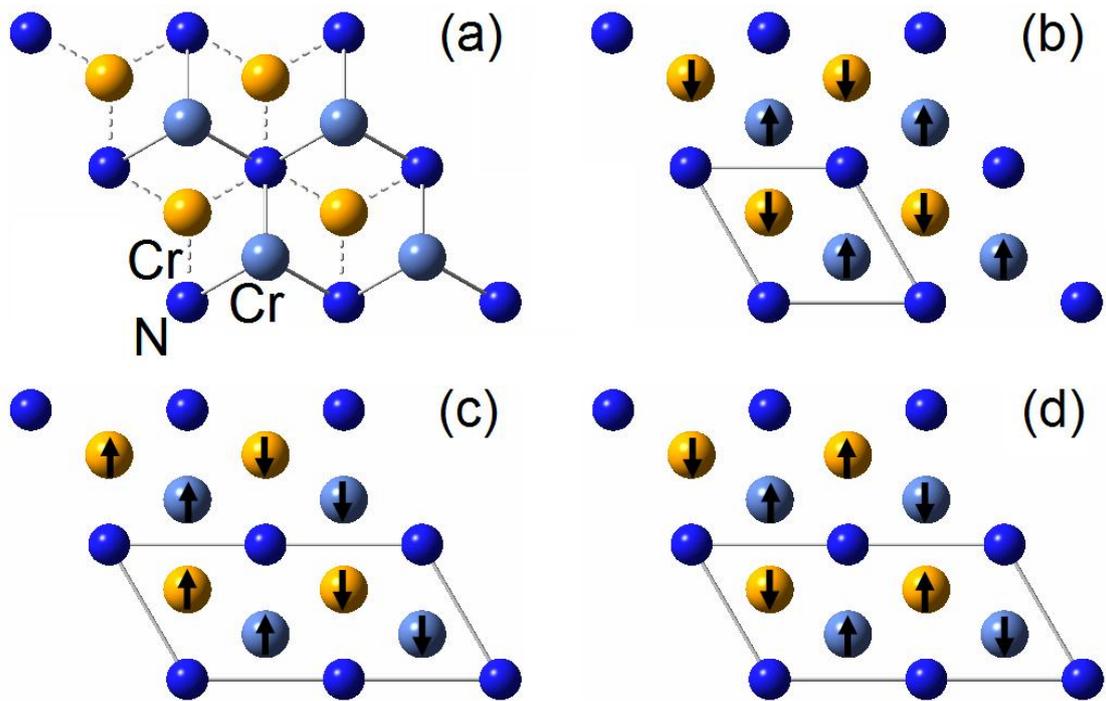

Figure 1. (a) Top view of two-dimensional Ni$_2$N, (b)−(d) are AFM1, AFM2, and AFM3 configurations. The adopted unit cells are denoted by tetragons. The Ni atoms in different layers are labeled with different colors.

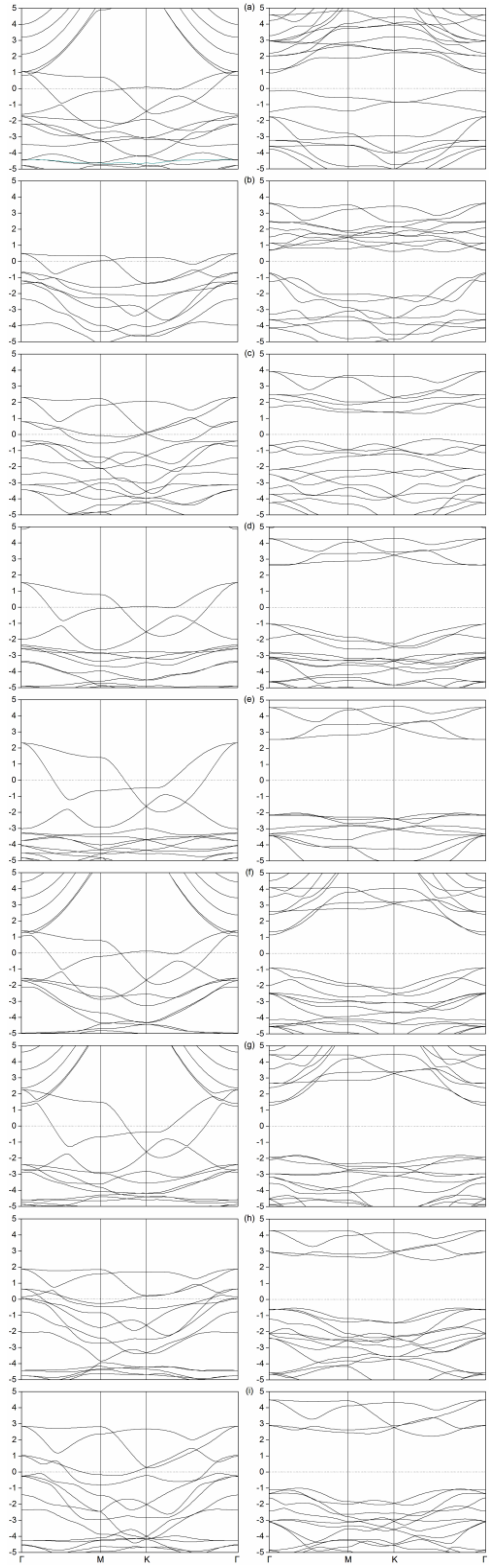

Figure 2. Band structures of (a) $Fe_2N(OH)_2$, (b) $Fe_2NO_2$, (c) $Co_2NO_2$, (d) $Ni_2NF_2$, (f) $Ni_2N(OH)_2$, and (h) $Ni_2NO_2$ with FM configurations, (e)−(i) are band structures of $Ni_2NF_2$, $Ni_2N(OH)_2$, and $Ni_2NO_2$ calculated with the HSE06 hybrid density functional. Vertical axis: energy in eV, horizontal axis: reciprocal lattice vector.

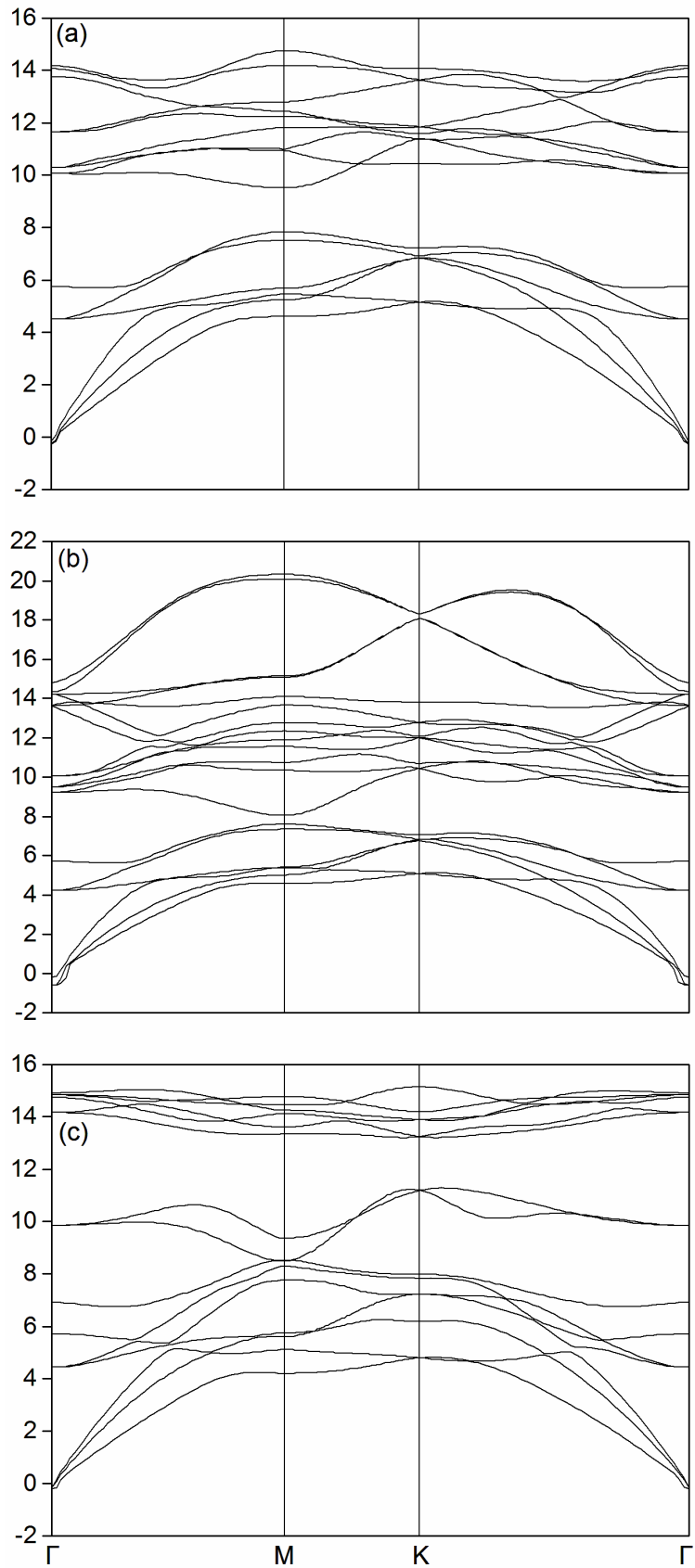

Figure 3. Phonon dispersion of (a) Ni$_2$NF$_2$, (b) Ni$_2$N(OH)$_2$, and (c) Ni$_2$NO$_2$. Vertical axis: frequency in THz, horizontal axis: reciprocal lattice vector.